# Two-Dimensional Geometry of Spin Excitations in the High Temperature Superconductor YBa$_2$Cu$_3$O$_{6+x}$

V. Hinkov*, S. Pailhès†, P. Bourges†, Y. Sidis†, A. Ivanov‡, A. Kulakov*, C.T. Lin*, D.P. Chen*, C. Bernhard* and B. Keimer*

* Max-Planck-Institut für Festkörperforschung, 70569 Stuttgart, Germany

† Laboratoire Léon Brillouin, CEA-CNRS, CE-Saclay, 91191 Gif-sur-Yvette, France

‡ Institut Laue-Langevin, 156X, 38042 Grenoble cedex 9, France

The fundamental building block of the copper oxide superconductors is a Cu$_4$O$_4$ square plaquette. In most of these materials, the plaquettes are slightly distorted and form a rectangular lattice. An influential theory predicts that high-temperature superconductivity is nucleated in "stripes" aligned along one of the axes of this lattice[1–3]. This theory had received strong support from experiments that appeared to indicate a one-dimensional character of the magnetic excitations in the high temperature superconductor YBa$_2$Cu$_3$O$_{6.6}$ (ref. 4). Here we report neutron scattering data on "untwinned" YBa$_2$Cu$_3$O$_{6+x}$ crystals, in which the orientation of the rectangular lattice is maintained throughout the entire volume. Contrary to the earlier claim[4], we demonstrate that the geometry of the magnetic fluctuations is *two-dimensional*. Rigid stripe arrays therefore appear to be ruled out over a wide range of doping levels in YBa$_2$Cu$_3$O$_{6+x}$, but the data may be consistent with liquid-crystalline stripe order[5]. The debate about stripe theories of high-temperature superconductivity has thus been reopened. In the future, it will be strongly constrained by detailed information about the in-plane anisotropy of the magnetic fluctuations revealed by our experiments.

Inelastic neutron scattering data can pinpoint the theoretically predicted stripe phases by virtue of their characteristic quasi-one-dimensional spin excitations[6]. Spin excitations incommensurate with the crystal lattice have indeed been observed by neutron scattering in several families of layered cuprates[4,7-12]. However, almost all neutron experiments thus far reported have been carried out on fully "twinned" crystals with equal proportions of micron-scale twin domains in which the rectangular $Cu_4O_4$ plaquettes are rotated by 90° with respect to one another[7-11]. Since the scattering pattern from such crystals consists of equal contributions from both twin domains, even perfectly one-dimensional spin fluctuations generate a four-fold symmetric pattern, so that they cannot be discriminated from microscopically two-dimensional fluctuations. The results of two neutron scattering experiments on partially detwinned $YBa_2Cu_3O_{6+x}$ crystals[4,12] have been interpreted as evidence of a one-dimensional character of the magnetic fluctuations[4,12-14]. However, due to significant contributions from the minority domain, the full geometry of the excitation spectrum has remained unclear.

Using neutron scattering from a mosaic of fully untwinned, nearly optimally doped $YBa_2Cu_3O_{6.85}$ crystals (see Methods) we have resolved this issue. We find that the locus of maximum spin fluctuation spectral weight approximately forms a circle in momentum space, so that the basic character of the spin excitations is two-dimensional. However, the damping and amplitude along the circle are modulated in a one-dimensional fashion. The strength of this modulation depends strongly on the excitation energy. Data from underdoped, untwinned $YBa_2Cu_3O_{6.6}$ crystals exhibit similar features.

Scans through the (200) and (020) Bragg reflections of the $YBa_2Cu_3O_{6.85}$ array reveal a bulk population ratio between majority and minority twin domains of about 95:5 (Fig. 1b and Methods). This is one order of magnitude larger than in previous experiments[4,12]. The contribution of the minority domain to the magnetic scattering pattern is thus negligible in our experiments.

Figs. 2 and 3 show magnetic neutron scattering data from this crystal array. The overall features of the neutron cross section are in good agreement with prior work on twinned crystals[7,9–11]. Due to antiferromagnetic correlations between spins in the $CuO_2$ plane, the spectral weight of low energy spin excitations in $YBa_2Cu_3O_{6+x}$ is centered around the in-plane wave vector $\mathbf{Q}_{AF} = (0.5, 1.5)$. As a consequence of antiferromagnetic interactions within a bilayer unit, the magnetic intensity is modulated as a function of the $c$-axis wave vector transfer, $L$, such that the maximum intensity is observed at odd multiples of $L = 1.7$. The characteristic $L$-dependence of the intensity implies that the signal originates from the $CuO_2$ planes. As observed before, the magnetic excitations are strongly enhanced below the superconducting transition temperature. For excitation energies below the "spin gap", $\hbar\omega_g = 30$ meV, the magnetic spectral weight is below the detection limit. The lowest-energy excitations observed above $\hbar\omega_g$ are incommensurate (Fig. 2). The incommensurate excitation branches disperse towards $\mathbf{Q}_{AF}$ with increasing excitation energy (Fig. 3), and they merge at $\hbar\omega_0 = 41$ meV giving rise to the "resonance peak".

The new aspect of this work is the determination of the in-plane geometry of the spin excitations. Fig. 2c–k shows representative scans from a comprehensive map of the spin fluctuation spectral weight at $\hbar\omega = 35$ meV. More limited data sets were also taken at $\hbar\omega = 31, 33, 37$ (not shown), and 38 meV (Fig. 2a,b). In order to extract the magnetic spectral weight from the experimentally determined scattering profiles, we have numerically convoluted an anisotropic damped harmonic oscillator cross-section with the spectrometer resolution function (see Supplementary Material). The computed profiles, shown as solid lines in Figs. 2 and 3, provide excellent descriptions of the experimental data. The intrinsic magnetic spectral weight at 35 meV extracted from this analysis is depicted in Fig. 1c.

We emphasize two salient features of the magnetic spectrum of $YBa_2Cu_3O_{6.85}$. The most important observation is that well-defined incommensurate peaks are present in scans along both $a^*$ and $b^*$ (Fig. 2a-d). This demonstrates the intrinsic *two-dimensional* nature of the spin excitations. The displacements from the commensurate wave vector $\mathbf{Q}_{AF}$ along $a^*$ and $b^*$ are identical within an experimental error of 15% at all energies covered by our experiment. The locus of maximum magnetic intensity approximately forms a circle around $\mathbf{Q}_{AF}$, following an *isotropic* dispersion relation $\omega_{\mathbf{Q}} = \omega_0 - c(\mathbf{Q} - \mathbf{Q}_{AF})^2$ with parameters $\hbar\omega_0 = 41 \pm 1$ meV and $\hbar c = (180 \pm 30)$ meVÅ$^2$ for all energies.

Second, both the amplitude and the width of the incommensurate peaks exhibit pronounced in-plane anisotropies (Fig. 1c), which are strongly energy dependent. At $\hbar\omega = 35$ meV, for instance, the scans along $a^*$ (Fig. 2c) are resolution limited, but those along $b^*$ (Fig. 2d) are much broader. However, the |Q|-integrated spectral weights along $a^*$ and $b^*$ differ by at most 40%. At $\hbar\omega = 38$ meV, the width anisotropy is significantly smaller than at 35 meV (Fig. 2a,b), and the |Q|-integrated intensities along $a^*$ and $b^*$ are identical to within 15%.

In order to check the generality of our findings, we have repeated our experiment on an array of $YBa_2Cu_3O_{6.6}$ crystals ($T_c = 61$ K) with a 94:6 detwinning ratio. Representative data are shown in Fig. 4. In accord with prior work on twinned crystals[7,9,12], the resonance peak is observed at a lower energy, $\hbar\omega_0 = 37.5$ meV, than at optimum doping. The salient features of the incommensurate excitations below $\hbar\omega_0$ are, however, quite similar. Specifically, incommensurate peaks are observed along both $a^*$ and $b^*$ at all energies monitored in our experiment, so that the geometry of the excitation spectrum is *two-dimensional*. The amplitude and width anisotropy is comparable to that at optimum doping, and it exhibits the same increase with decreasing energy. Even at 28 meV, i.e. 10 meV below the resonance peak, the |Q|-integrated spectral weights along a* and b* differ by at most 50%. The only significant difference is that the dispersion relation is somewhat steeper along *b** than along *a**, whereas it is isotropic at optimum doping.

An anisotropy of the magnetic signal along *a** and *b** was first pointed out by Mook et al.[4] for a $YBa_2Cu_3O_{6.6}$ crystal with a detwinning ratio of 2:1. From a limited data set at a single energy, they deduced a 100% intensity difference between the two directions and interpreted their data as evidence of a one-dimensional spin fluctuation geometry. Based on an analysis of a large set of data between $\hbar\omega_g$ and $\hbar\omega_0$, we have now, however, demonstrated that the low-energy spin excitations of both underdoped and optimally doped $YBa_2Cu_3O_{6+x}$ have a two-dimensional geometry. Which models can describe the observed one-dimensional amplitude and width anisotropy?

Theories based on a one-dimensional, rigid array of stripes predict a 100% intensity anisotropy and cannot account for the two-dimensional scattering pattern[13,14]. The map of the magnetic intensity at 35 meV does, however, bear resemblance to the scattering pattern generated by a nematic liquid crystal close to a nematic-to-smectic critical point. In this scenario, the structural anisotropy between *a*- and *b*-axes may act as an aligning field for the nematic director[6]. A preferential alignment of charge stripes along *b* would enhance the magnetic spectral weight along *a*. However, detailed theoretical work is required to ascertain whether the observed scattering pattern is indeed a unique signature of a quantum fluid of fluctuating stripes. A theoretical description in this framework should also account for the observed energy dependence of the anisotropy. An alternative approach would be to consider a two-dimensional arrangement of charged domain walls[1,15].

In the absence of experimental information about the in-plane anisotropy, prior Fermi liquid-based theoretical scenarios for the spin dynamics of $YBa_2Cu_3O_{6+x}$ had only considered the influence of the $CuO_2$ layers, ignoring a possible influence of the *b*-axis oriented CuO chains[16,17]. Based on the observed *c*-axis modulation of the intensity, we can rule out a direct contribution from the CuO chains to the observed magnetic signal. However, an indirect influence of the CuO chains on the spin dynamics in the layers cannot be excluded. The plane-derived Fermi surfaces are predicted to exhibit a small in-plane anisotropy due to the hybridization between orbitals centered on chains and layers and the different in-plane Cu-O bond lengths along *a* and *b*[18]. Further, Nuclear Quadrupole Resonance experiments[19] show that charge density waves in the CuO chains may induce a charge modulation in the $CuO_2$ layers, and X-ray scattering experiments[20] indicate a polarization of Cu orbitals in the $CuO_2$ layers due to oxygen ordering in the chain layers. The impact of these factors on the spin dynamics will have to be assessed in quantitative calculations. Other factors, such as proximity to a "Pomeranchuk" instability of the Fermi surface[21–23], may also contribute to the anisotropy of the spin dynamics. The observed in-plane anisotropy of the superconducting energy gap imposes further constraints on theory[24,25].

In conclusion, our data put stringent, quantitative constraints on stripe theories of high temperature superconductivity. They rule out a large class of theories according to which high-$T_c$ superconductivity coexists with a rigid array of stripes, over a wide range of hole concentrations encompassing the 60K and 90K superconducting phases of $YBa_2Cu_3O_{6+x}$. If interpreted as evidence of the theoretically predicted liquid-crystalline stripe phases, they quantify the orientational stripe fluctuations.

## Methods

**Sample preparation** High-quality, single phase $YBa_2Cu_3O_{6+x}$ crystals were synthesized by the solution growth method and cut into rectangular shapes. For the nearly optimally doped sample, 82 crystals with a total mass of ~ 1.3 g were annealed at 520°C in $O_2$ resulting in an oxygen content of x ~ 0.85. The crystals were individually detwinned by subjecting them to a uniaxial mechanical stress of ~ $5 \times 10^7$ N/m$^2$ along the <100> axis at 520°C in flowing oxygen for at least 4 hours[26,27]. The resulting specimens show superconducting transition temperatures, $T_c$, of 90 K with $\Delta T_c$ between 1 K and 3 K. They were then co-aligned on two Al-plates into which a dense grid of holes had been drilled. By using thin Al wire to fix the crystals to the grid, it proved possible to largely avoid organic glue, and hence to substantially reduce the background signal due to incoherent neutron scattering from hydrogen. The full array shows a mosaicity of < 1.2° and is almost entirely untwinned (Fig. 1b). The underdoped sample consists of 120 crystals with a total mass of ~ 2 g. After annealing at 590°C in air and detwinning at 400°C in Ar, these specimens show a $T_c$ of 61 K with $\Delta T_c$ between 1 K and 2 K. They were co-aligned on two Al-plates and fixed by Al-screws, thus completely avoiding organic glue. The mosaicity is ~ 1.2° and the domain population ratio 94:6.

**Neutron scattering** The wave vector $\mathbf{Q} = (H, K, L)$ is quoted in units of the reciprocal lattice vectors $a^*$, $b^*$ and $c^*$, where $a = 2\pi/a^* = 3.82$ Å, $b = 2\pi/b^* = 3.88$ Å, and $c = 2\pi/c^* = 11.7$ Å. Because of the slightly different lengths of the lattice parameters $a$ and $b$ in the $CuO_2$ planes, the (200) and (020) crystallographic Bragg reflections occur at different wave vector transfers $\mathbf{Q}$. Since neutron scattering is a bulk probe, scans through these reflections directly reveal the bulk population ratio of the two twin domains, as shown in Fig. 1b.

The data for nearly optimally doped $YBa_2Cu_3O_{6.85}$ were collected on the 2T triple-axis spectrometer at the Laboratoire Léon Brillouin (LLB), Saclay (France). Because of kinematical restrictions, the scans were carried out in the second Brillouin zone (Fig. 1A). A specially designed holder was used, which enabled us to access the $P_+$ = {(1, 0, 0) ; (0, 3, 3.4)} scattering plane as well as the complementary $P_-$ = {(0, −1, 0) ; (3, 0, 3.4)} plane during the experiment without opening the cooling chamber. This change between $P_+$ and $P_-$ is equivalent to a rotation of the resolution ellipsoid by 90° with respect to $a$* and $b$* and allows a quick check for resolution effects. No collimators were used in order to maximize the neutron flux. Graphite filters extinguished higher order contamination of the neutron beam. The data for underdoped $YBa_2Cu_3O_{6.6}$ were collected on the IN8 spectrometer at the Institut Laue-Langevin, Grenoble (France). The same holder was employed and the spectrometer was configured in the same way. The final wave vector was fixed to either 2.662 Å$^{-1}$ (Figs. 2, 3c-f, 4) or 3.85 Å$^{-1}$ (Fig. 3a,b).

**Data analysis** Since the magnetic excitations are overdamped at high temperatures, the spin fluctuation contribution to the neutron cross section is most clearly apparent in temperature subtractions. For $YBa_2Cu_3O_{6.85}$ we show subtractions between the intensities at T = 10 K and T = 100 K. For $YBa_2Cu_3O_{6.6}$, which exhibits a significant magnetic signal in the normal state above $T_c$ = 61 K, we show subtractions between T = 5 K and T = 250 K. The simplest model consistent with our data is an anisotropic damped harmonic oscillator with an isotropic dispersion law, isotropic distribution of the |Q|-integrated intensity, and anisotropic damping parameter in the $a$*/$b$*-plane (see Supplementary Information). The computed profiles shown in Figs. 2 and 3 are numerical convolutions of this model with the spectrometer resolution function. The numerical resolution convolution was performed with the program Rescal-5 based on the Popovici method[28] (see online description under http://www.ill.fr/tas/matlab/doc/rescal5/rescal.htm) as well as with in-house programs at LLB.

**Supplementary Information** accompanies the paper on **www.nature.com/nature**.

**Acknowledgements** We thank B. Hennion, S. Kivelson, D. Manske, W. Metzner and H. Yamase for discussions, S. Lacher, H. Wendel, B. Baum and M. Bakr for crystal preparation, M. Ohl and W. Plenert for the design and manufacturing of the sample holder, H. Bender, C. Busch and H. Klann for technical support, and P. Baroni for technical support at the 2T spectrometer. We acknowledge support from the

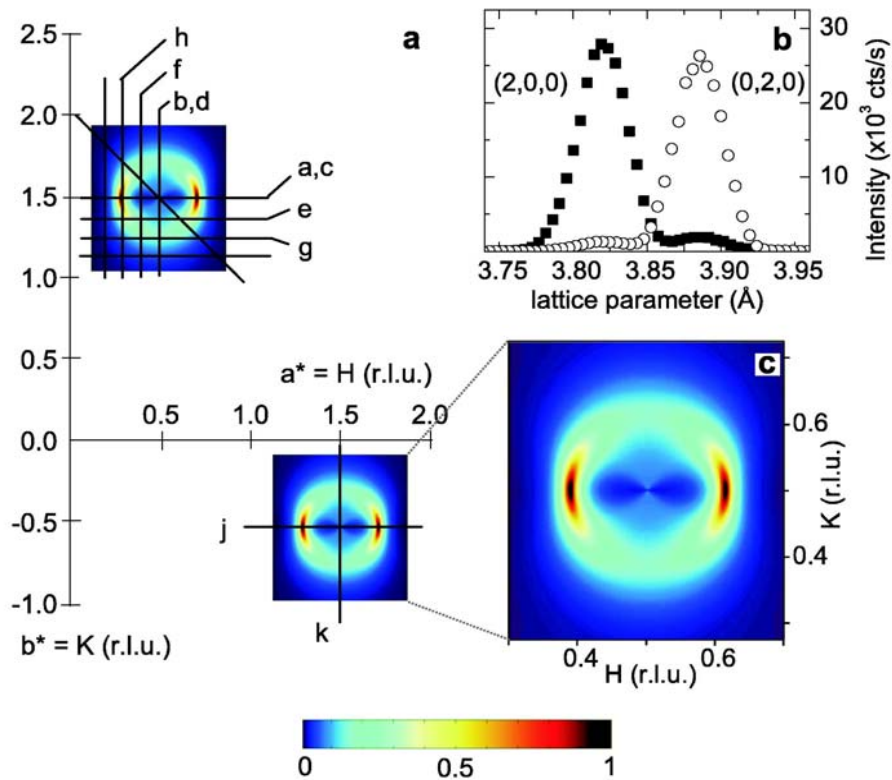

**Figure 1** Layout of the reciprocal lattice and magnetic spectral weight of YBa$_2$Cu$_3$O$_{6.85}$. **a**, In-plane projection of the YBa$_2$Cu$_3$O$_{6.85}$ reciprocal lattice indicating the trajectories of the constant-energy scans shown in Fig. 2. The representation is not true-to-scale. **b**, Longitudinal elastic scans through the (2, 0, 0) and (0, 2, 0) crystallographic Bragg reflections, demonstrating a twin domain population ratio of ~ 95:5. **c**, Intrinsic magnetic spectral weight at 35 meV extracted from the analysis presented in the supplementary material.

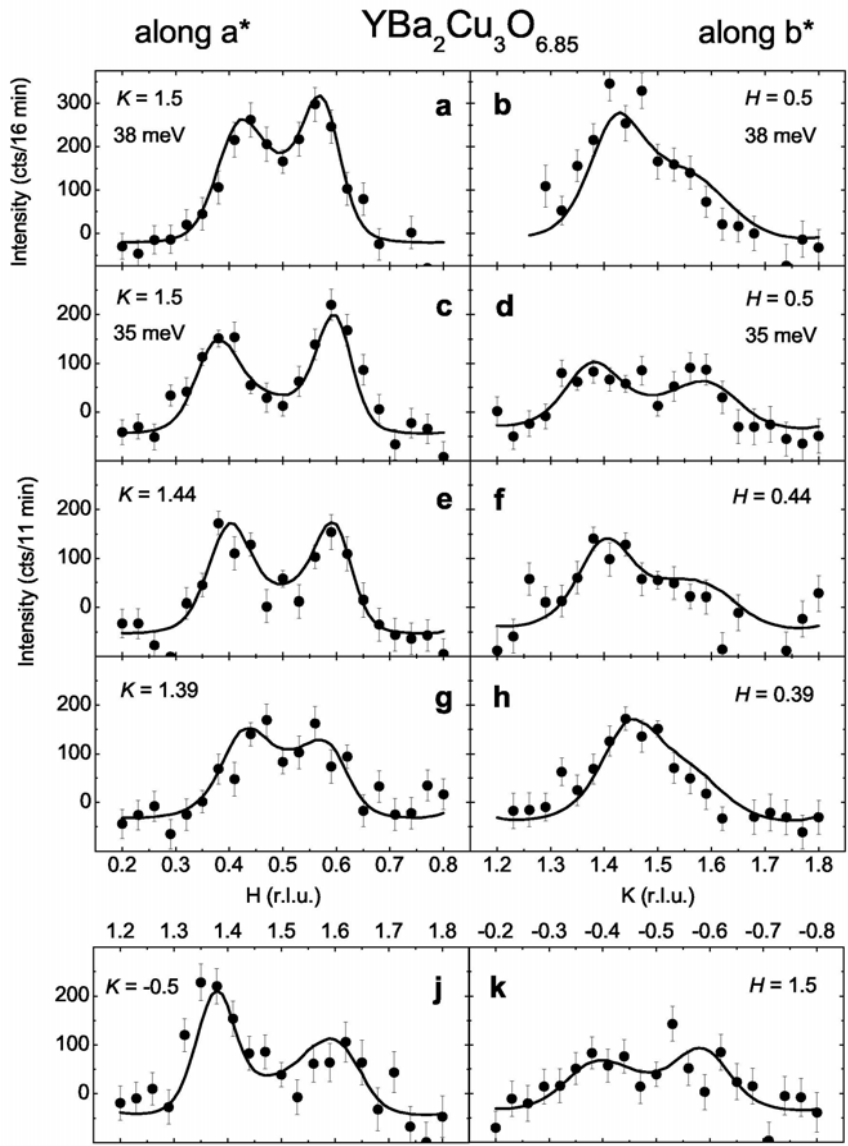

**Figure 2** Constant-energy scans along the trajectories indicated in Fig. 1a. The excitation energies are $\hbar\omega$ = 38 meV (**a**,**b**) and 35 meV (**c** − **k**), and **Q** = (*H, K*,1.7). We show subtractions of the intensities at T=10 K (< $T_c$) and T=100 K (> $T_c$). The lines result from a convolution of a model cross section with the instrumental resolution function (see Methods). The data in panels **c**,**d** and **j**,**k** were taken in two different Brillouin zones with exchanged resolution conditions. The observed anisotropy between a* and b* is thus not due to resolution effects.

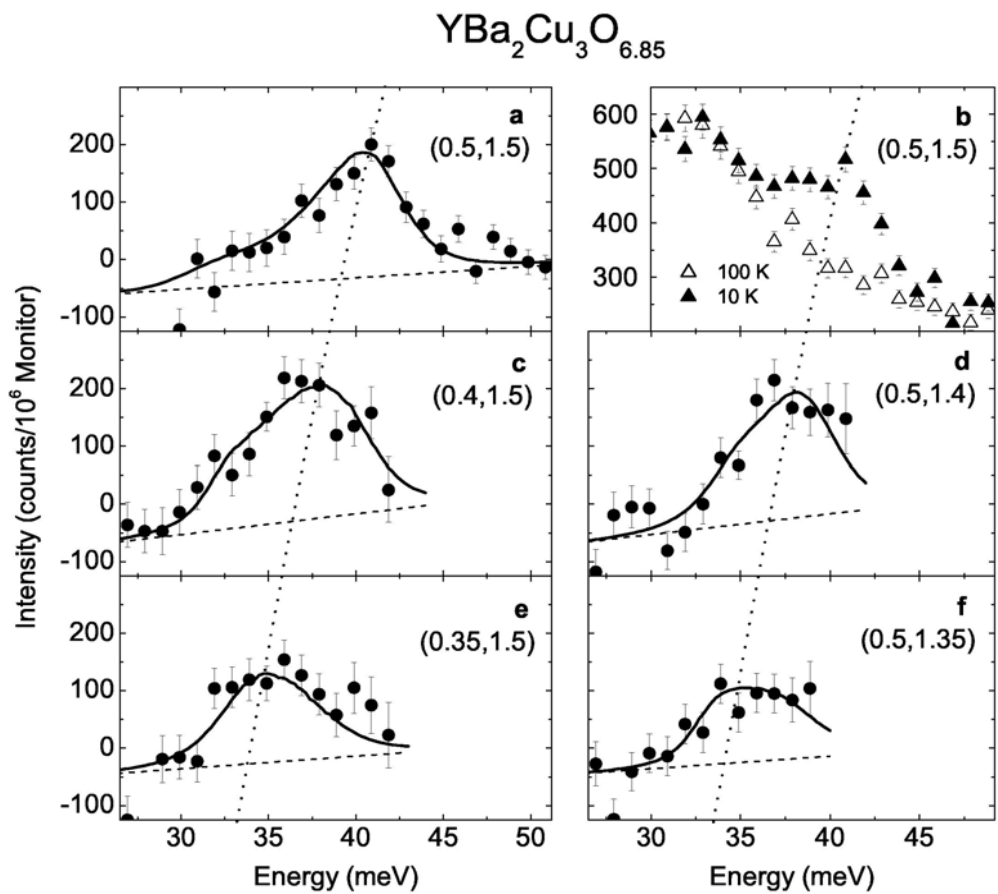

**Figure 3** Constant-**Q** scans demonstrating the dispersion of the magnetic mode along *a*\* (left) and *b*\* (right). Panel **b** shows raw data above and below T$_c$. The remaining panels show subtractions of the intensities at T=10 K (<T$_c$) and T=100 K (>T$_c$). The solid lines result from a numerical convolution of a model cross section with the instrumental resolution function (see Methods). The dashed lines show the slightly energy dependent background due to thermal population of phonon excitations.

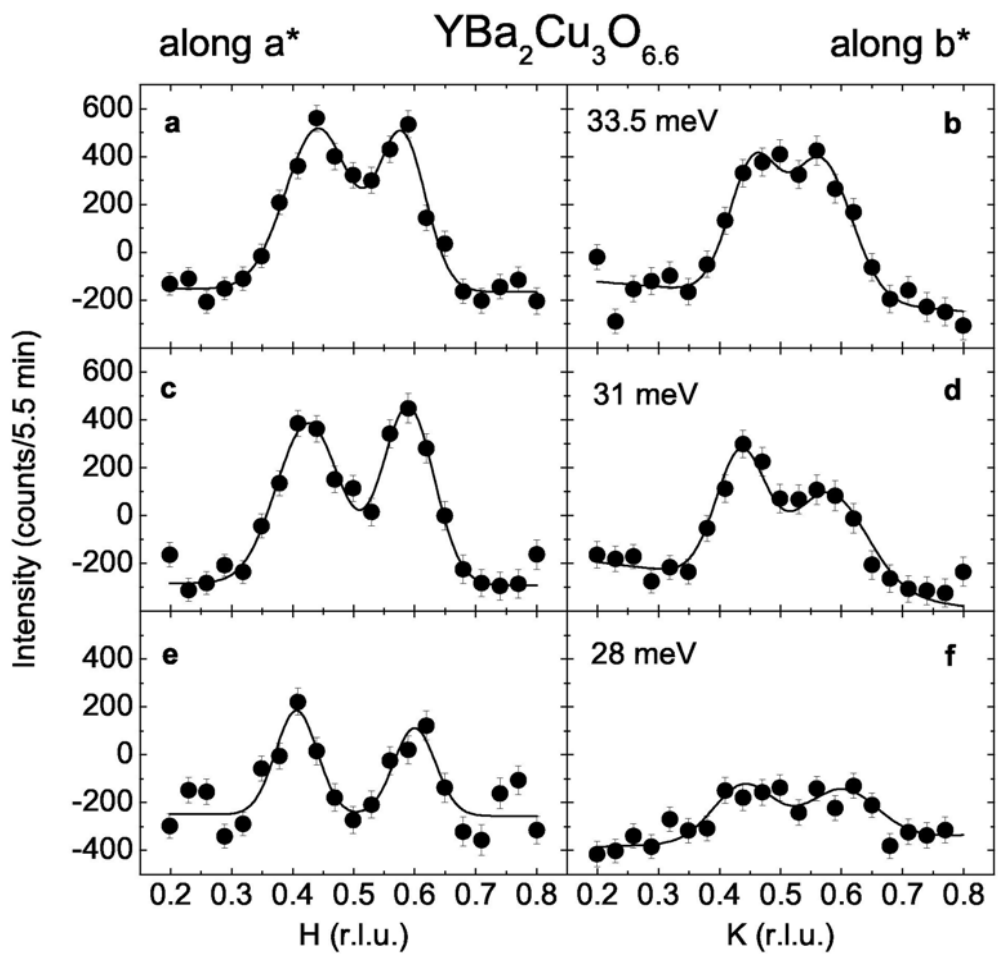

**Figure 4** Constant-energy scans for YBa₂Cu₃O₆.₆. They were performed along *a*\* at **Q** = (H, −1.5, 1.7) (left) and along *b*\* at **Q** = (1.5, K, 1.7) (right), at different energies ℏω. The plots show subtractions of the intensities at T = 5 K and T = 250 K. The lines are double Gaussians fitted to the data. The resolution conditions were identical for measurements along *a*\* and *b*\*, so that the scans are directly comparable.

**Supplementary Material**

**Analysis of Inelastic Neutron Scattering Data of Detwinned YBa$_2$Cu$_3$O$_{6.85}$ Within an Anisotropic, Damped Harmonic Oscillator Model**

In order to extract the magnetic spectral weight from the experimentally determined scattering profiles, we have numerically convoluted an anisotropic damped harmonic oscillator cross section

$$I(\mathbf{Q}, \omega) = I_{\mathbf{Q}}^0 \frac{\omega \gamma_{\mathbf{Q}}}{\left(\omega^2 - \omega_{\mathbf{Q}}^2\right)^2 + \left(\omega \gamma_{\mathbf{Q}}\right)^2} \qquad (1)$$

with the spectrometer resolution function. An *isotropic* dispersion $\omega_{\mathbf{Q}} = \omega_0 - c(\mathbf{Q} - \mathbf{Q}_{AF})^2$ with parameters $\hbar\omega_0$ 41 ± 1 meV and $\hbar c = (180 \pm 30)$ meVÅ$^2$ accounts for the locus of incommensurate peaks in momentum space for all energies, in agreement with previous reports for similar doping levels[1,2]. The observed anisotropy of the amplitude and width can be parameterised as $I_{\mathbf{Q}}^0 = I_0 + \Delta I(\omega)(1 - \cos(2\varphi_{\mathbf{Q}}))/2$ and $\gamma_{\mathbf{Q}} = \gamma_0 + \Delta\gamma(\omega)(1 - \cos(2\varphi_{\mathbf{Q}}))/2$, respectively, where $\varphi_{\mathbf{Q}} = \arctan\left[(K - K_{AF})/(H - H_{AF})\right]$ is the azimuthal angle in the CuO$_2$ layer.

The constant-energy cuts along $a^*$ ($\varphi_{\mathbf{Q}} = 0$) are resolution limited within the error, so that we can only establish an upper bound of 3 meV on the damping parameter $\gamma_0$. In contrast, the profiles along $b^*$ are significantly broader than the experimental resolution for 35 meV but become sharper with increasing energy (Fig. 2b,d, main text). In the simplest model that agrees with all of our data, the |Q|-integrated intensity is independent of $\varphi_{\mathbf{Q}}$, i.e. $\Delta I = 0$ for all energies. The solid lines in Figs. 2 and 3 are results of a numerical convolution of this model with the spectrometer resolution function. The good agreement of the computed profiles with the data confirms the intrinsic nature of the observed anisotropy. Fig. 1c provides a full representation of the intrinsic magnetic spectral weight at 35 meV resulting from this analysis. The energy evolution of the damping parameter $\gamma$ along $a^*$ and $b^*$ is shown in Fig. S1. The confidence intervals for $\gamma$ were established by systematically fitting the data using model cross sections with different anisotropy parameters $\Delta\gamma$ and $\Delta I$. This procedure also yields a lower bound on $\Delta I$ of $\Delta I \geq -0.4$ at 35 meV and $\Delta I \geq -0.15$ at 38 meV.

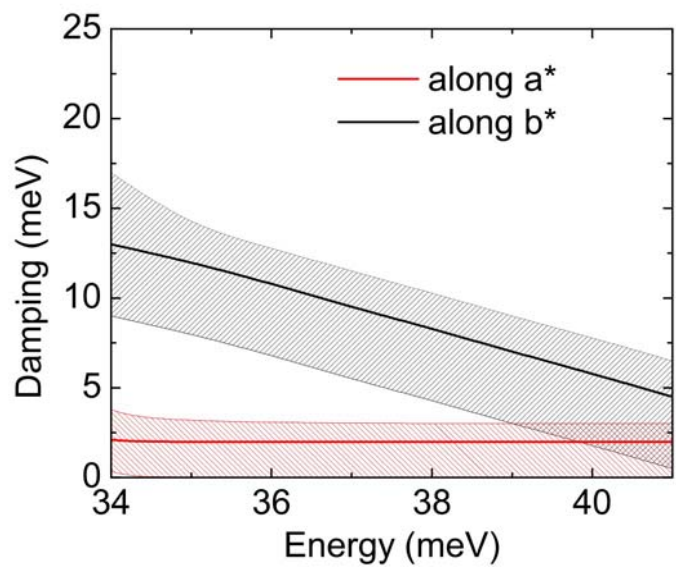

**Figure S1** Energy dependence of the damping parameter along $a^*$ and $b^*$. The shaded area indicates the corresponding confidence intervals.